# (Sub)nanoscale Visualization of Reconstruction-Driven Moiré Exciton Localization and Delocalization

Medha Dandu[1,^,#], Sriram Sankar[2,#], Giovanny Espitia[3#], Rwik Dutta[3], Patrick J. Hays[2], Daria Blach[1], Takashi Taniguchi[4], Kenji Watanabe[4], James R. Chelikowsky[3,6,7], Seth Ariel Tongay[2], Peter Ercius[1], Jordan A. Hachtel[5], Mit H. Naik[3,6*], Archana Raja[1,8*], Sandhya Susarla[2,*]

#- Equal contribution
*Corresponding author
1. Molecular Foundry, Lawrence Berkeley National Laboratory, USA, 94720
2. Materials Science and Engineering, School for Engineering of Matter, Transport and Energy, Arizona State University, USA,85287
3. Department of Physics, University of Texas at Austin, Austin, TX, USA, 78712
4. National Institute for Materials Science, Japan, 3050044
5. Center for Nanophase Materials Sciences, Oak Ridge National Laboratory, TN, USA,37830
6. Oden Institute for Computational Engineering and Sciences, The University of Texas at Austin, Austin, Texas 78712, USA
7. McKetta Department of Chemical Engineering, The University of Texas at Austin, Austin, Texas 78712, USA
8. Kavli Energy NanoScience Institute, University of California Berkeley, Berkeley, CA, USA, 94720
^ Current affiliation: Jawaharlal Nehru Centre for Advanced Scientific Research, Bangalore, India, 560064

## Abstract:

Spectral fingerprints in optical absorption and emission have typically been used as a signature of exciton localization in twisted moiré bilayers. However, the mechanism by which excitons become confined to specific stacking sites and their associated optical signature is experimentally unresolved. Here, we directly visualize in real space how tuning the change in extent of structural reconstruction leads to localization and delocalization of moiré excitons in the $WSe_2/WS_2$ moiré superlattice. Using cryogenic monochromated electron energy loss spectroscopy, together with first-principles GW-Bethe Salpeter equation calculations and optical spectroscopy, we uncover the physical mechanism that drives the correlation between twist-angle-dependent structural transformations, the real-space localization of moiré excitons, and their optical signatures. Surprisingly, and in contrast to the prevailing understanding, we show that the emergence of new moiré exciton resonances in the optical spectra alone is insufficient to establish exciton localization. Instead, the extent of structural reconstruction and external strain drives exciton localization, leading to new design principles for engineering moiré excitons and strain-aware quantum optoelectronic devices.

## Introduction

Twisting two transition metal dichalcogenides with respect to each other, complemented by tuning knobs such as strain[1], offers a versatile platform to engineer spatially modulated excitonic states [2–8]. These are electron-hole pairs bound by Coulomb interactions that strongly interact with the moiré potential and are localized into reconfigurable and correlated quantum confined states [2–7,9–14]. Exciton localization in a semiconducting moiré heterostructure is indirectly identified in reflectance spectroscopy by the splitting of the lowest-lying exciton peak [10,12,15–21]. The authors[13,22], along with others, have identified that a key driver of exciton confinement is structural relaxation within the moiré unit cell, where individual stacking sites undergo twist-dependent reconstruction [23–29]. This can flatten the electronic bands and localize excitons at specific regions of the unit cell [13,29–33]. Such confinement can be further perturbed by nanofabrication-induced inhomogeneities, including nanobubbles and bend contours.[34,35]. Despite this progress, a unified picture connecting structural reconstruction, external strain, and exciton behavior in moiré heterostructures has yet to emerge. Resolving these correlations is critical not only for understanding the microscopic origins of exciton confinement but also for enabling deliberate, strain-mediated control over excitonic states in next-generation moiré devices.

Resolving these correlations through conventional far-field and near-field optical microscopy and spectroscopy[4,12,30,36–39] is challenging due to the lack of sub-nanometer spatial resolution required for imaging within a moiré unit-cell. This problem is equally formidable from a theoretical standpoint, as most computational models of the large moiré super-cells do not account for the effects of external strain [13,29]. To resolve this issue, we need a metrology that probes excitons with high spectral and spatial resolution. Monochromated scanning transmission electron microscopy electron energy loss spectroscopy (STEM-EELS) offers the unique capability to measure both spectral and spatial information of individual bright excitons simultaneously (**See Fig.1a**) by utilizing inelastic scattering [40]. While this technique has previously been employed by the authors[22] to image moiré excitons, insufficient spectral resolution and unit-cell averaging over a large area masked the effects of heterogeneous structural reconstruction and external strain.

In this study, we investigate how structural reconstruction and external strain influence exciton localization in $WS_2/WSe_2$ heterobilayers using cryogenic monochromated STEM-EELS with high spectral resolution, optical reflectance, and first-principles GW-Bethe Salpeter equation calculations (GW-BSE). We find that the relative intensities of moiré excitonic spectral features strongly correlate with the extent of structural reconstruction. We show that the external strain modifies the real-space location of exciton localization. Importantly, we demonstrate that the presence of additional excitonic peaks in optical reflectance alone does not constitute evidence of moiré-reconstruction driven exciton localization, unlike earlier predictions[29]. Instead, relative peak intensities and relative stacking-site areas prove to be the critical observables. We delineate a phase-space window for robust exciton localization, yielding design principles that inform moiré exciton experiments and establish a foundation for devices exploiting programmable exciton confinement.

## Results and Discussion

In an ideal $0^0$ aligned $WSe_2/WS_2$ bilayer, $WS_2$ and $WSe_2$ have a lattice mismatch of 4% with a type II band alignment, resulting in a moiré periodicity of ~8 nm with three high-symmetry stacking configurations: AA, $B^{Se/W}$, and $B^{W/S}$. At the AA stacking site, both the metal and chalcogen atomic columns are vertically aligned. At the Bernal stacking sites, $B^{Se/W}$ and $B^{W/S}$, the chalcogen atoms vertically align with the metal atoms. AA stacking has the largest interlayer separation and stacking energy due to the steric hindrance from the overlap of Se with S atoms from the $WSe_2$ and $WS_2$ layers. Under the absence of external strain, the superlattice undergoes a structural reconstruction to expand the regions of low-energy stacking while reducing the area of the high-energy AA stacking. This is achieved through local lattice distortions in which the lattice constant of $WSe_2$ is slightly reduced, and that of $WS_2$ is increased within the Bernal stacking regions. The displacement of the W atoms from our theoretical calculations in each layer relative to their unrelaxed positions is shown for the $WS_2$ and $WSe_2$ layers in **Fig 1b.** These displacements lead to the formation of a strain map characterized by alternating domains of compressive and tensile strain in each layer, which in turn modulate the electronic structure[37]. When two-dimensional materials are transferred onto TEM grids, external strain due to the presence of nanobubbles over a large field of view also plays a significant role (**See Fig S6)**. This type of strain is mostly tensile.

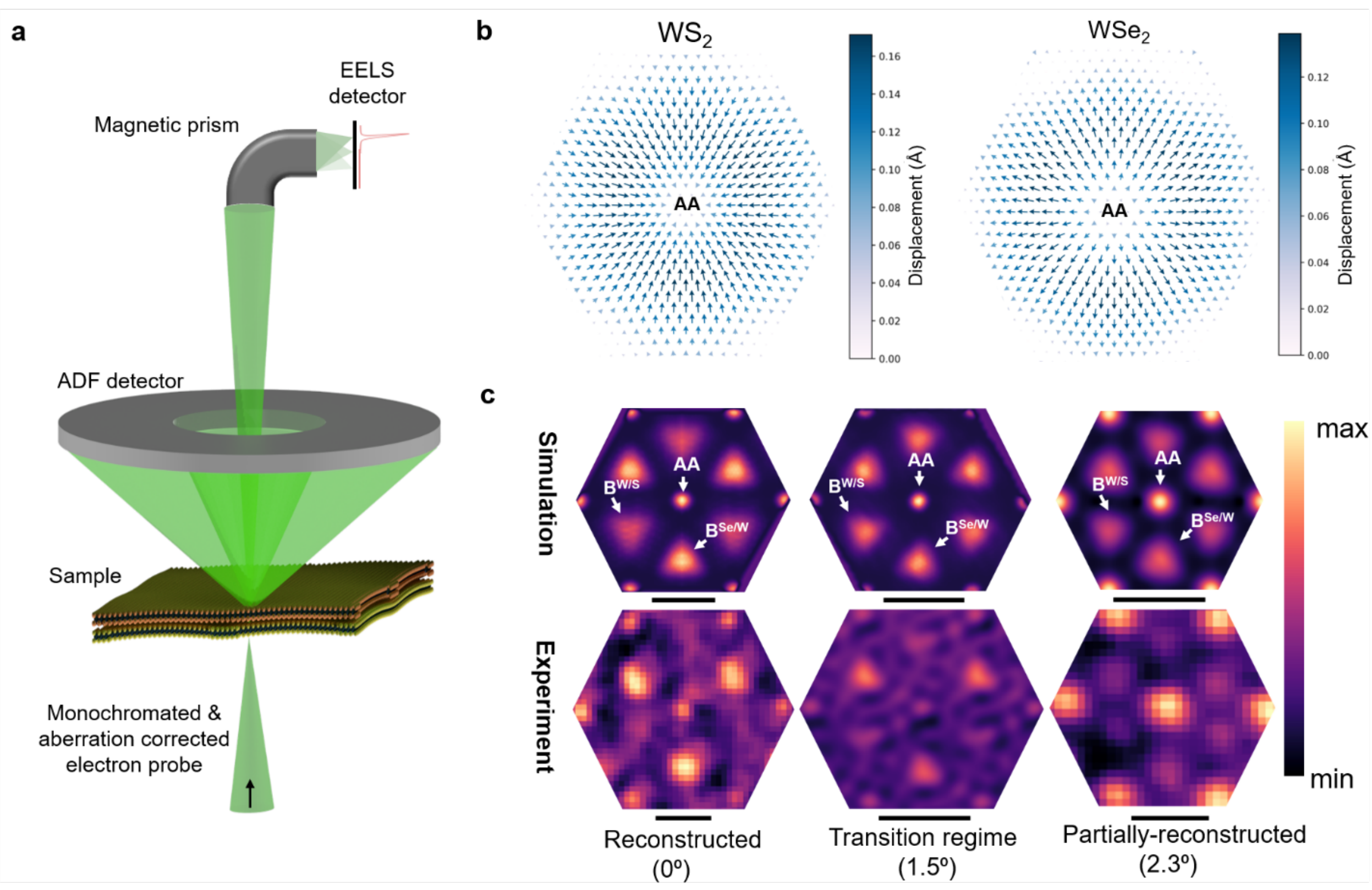

***Figure 1: Structural reconstruction in twisted $WSe_2/WS_2$.*** *(a) Schematic of the low-loss STEM-EELS technique. (b) Structural transformation around the AA stacking site in* $0^0$ *aligned $WSe_2/WS_2$*

*heterostructures. The arrows and colormap show the in-plane displacement of atoms from their unrelaxed to relaxed positions. (c) Simulated (top panel) and experimental (bottom panel) ADF-STEM images of different twisted $WSe_2/WS_2$ bilayers across varying reconstruction regimes. Simulated images use the relaxed atomic positions from our theoretical calculations. The scale bar is 5 nm.*

Experimentally, ADF-STEM cannot measure the strain due to reconstruction, but it can estimate the qualitative change in reconstruction by observing the relative size of the AA and Bernal stacking. On the other hand, the external strain can be estimated by comparing the lattice parameter of each layer over a large field of view. While the reconstruction is a combination of tensile and compressive strain at the angstrom scale level, the extrinsic strain is mostly tensile over nanoscale regimes due to the presence of nanobubbles and bend contours (**See Fig S6).** From here onwards, whenever strain is mentioned, it is mostly referring to external strain unless stated otherwise.

Experimentally, the large area ADF-STEM imaging was performed on different twisted heterostructures encapsulated with hexagonal boron nitride (hBN) (**Fig S1**) and pre-characterized for twist angles by polarization-resolved second harmonic generation (**Fig S2**). We directly observe the change in the extent of reconstruction with variation in the twist angle in experimental and computational HAADF-STEM images (**Fig 1c**). As the twist angle increases, the extent of structural reconstruction decreases gradually, manifesting as a systematic increase in the relative size of AA-stacked domains until the areas of AA, $B^{Se/W}$ and $B^{W/S}$ are roughly the same (**Fig 1c**), similar to earlier reports on twisted homobilayers[41,42]. Computational modeling corroborates this picture, revealing that differential intralayer strain at the individual stacking sites also reduces from a maximum value of ~1% to ~0.7% as the twist angle changes from 0° to 2.3° (**Fig S3**), accompanied by a corresponding suppression of out-of-plane corrugation. We classify this change into three regimes: reconstructed (0°), transition (1.5°), and partially-reconstructed (2.3°). Further, we also observe the effect of external strain by carefully observing the moiré unit cell. For the reconstructed regime, the computational image shows a perfect moiré unit cell whereas the experimental image shows a distorted moiré unit cell, indicating a strong presence of external strain.

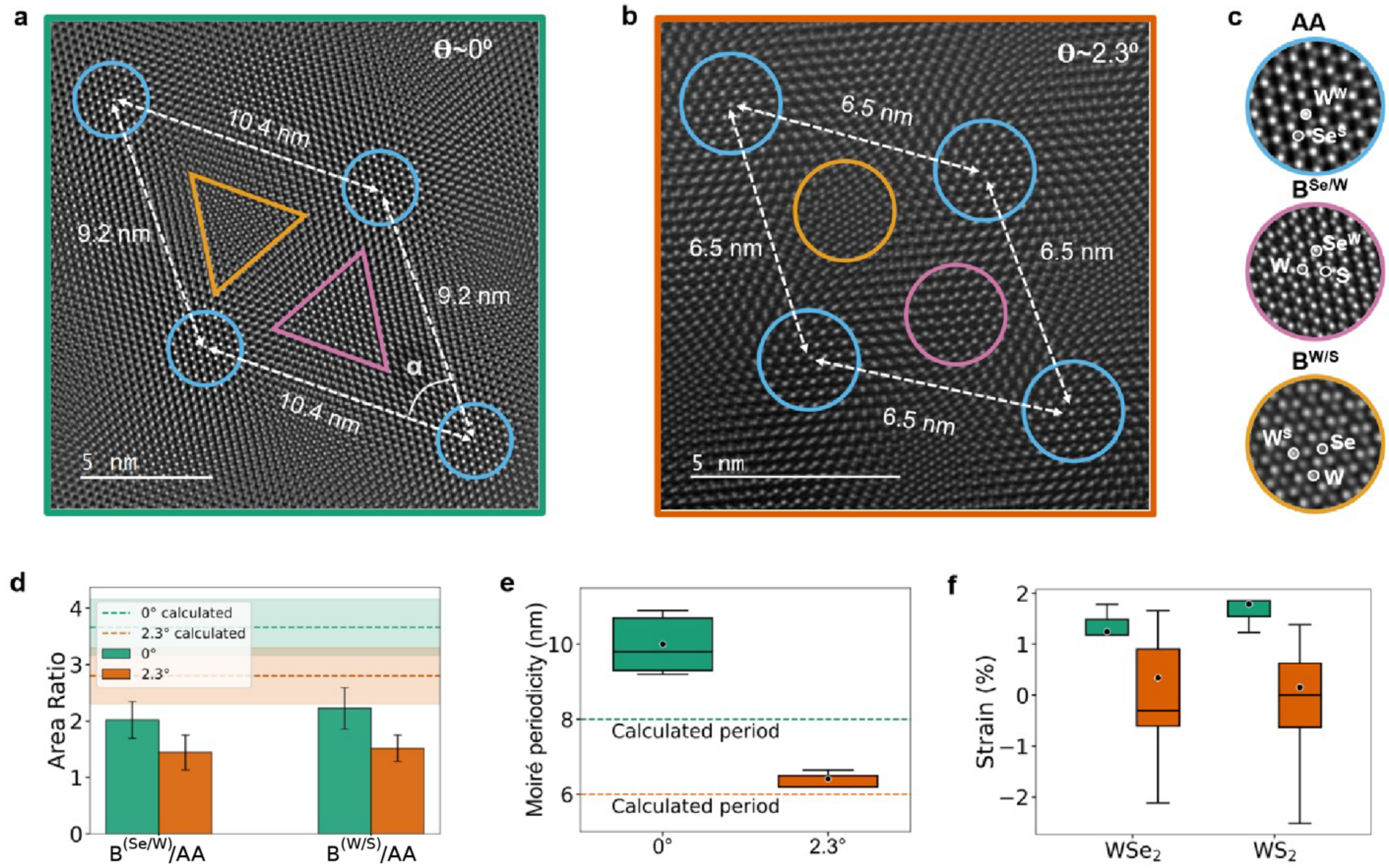

***Figure 2: Structural reconstruction and external strain effects.*** *High-resolution filtered ADF-STEM images from (a) reconstructed (green box) and (b) partially-reconstructed (orange box) $WSe_2/WS_2$ heterobilayers indicating different stacking registries – AA (blue), $B^{Se/W}$ (pink), and $B^{W/S}$(yellow). White dotted lines outline a single moiré unit cell. The scale bar is 5 nm. (c) Atomic configurations of AA, and Bernal ($B^{Se/W}$, and $B^{W/S}$) stackings. (d) Bernal ($B^{Se/W}$ and $B^{W/S}$) to AA stacking area ratio in reconstructed (green) and partially-reconstructed (orange) heterostructures, signifying the extent of structural reconstruction. (e) Estimated moiré periodicity from the experimental ADF-STEM images, signifying the external strain. Dotted lines in (d) and (e) represent the theoretically calculated area ratios and moiré periodicity from the multislice ADF-STEM simulations. (f) Layer-resolved strain from image filtering in the ADF-STEM images.*

To quantify the reconstruction and external strain better, we perform atomic resolution imaging as shown in **Fig 2 (a-b)** (details in **Fig S4**)**.** At first glance, we observe that AA sites are circular in shape in both the reconstructed and partially-reconstructed heterostructures. The Bernal stacking sites, however, form triangular domains in the reconstructed heterostructure and remain circular in the partially-reconstructed case. We further quantified the reconstruction using the area ratio of the Bernal and AA stacking sites. The area ratio changes from $2.2 \pm 0.5$ to $1.5 \pm 0.5$ as we transition from a reconstructed to a partially-reconstructed regime (**Fig 2d, details in Fig S5**). The multislice simulations performed on similar experimental conditions overestimate the area ratio (dotted lines in **Fig 2d**). This discrepancy may stem from varying hBN thickness, which changes the local contrast and the area ratio (**Fig S9,10**).

The reconstruction can further be perturbed by the external strain. We use moiré periodicity to gauge the effect of the external strain (**Fig 2e**). The moiré periodicity in the partially-reconstructed lattice (6.4 nm ± 0.05 nm) is similar to theoretical expectations. However, in a reconstructed lattice, the experimentally observed moiré periodicity increases to 10nm ± 0.2 nm, beyond the theoretical limit of 8 nm (**Fig 2e, Fig S5**). We corroborate this observation with the presence of nanobubbles in the lattice (**Fig S6**). We note that any local tilts in the 2D materials can result in "pseudo-strain"[43]. However, in such cases, the pseudo-strain is compressive, unlike the tensile strain that we observe from our current experiments (**Fig S11**). We estimated the heterogeneous external tensile strain by Fourier filtering of $WSe_2$ and $WS_2$ lattice spots and comparing it with the theoretically expected lattice parameters of each monolayer (Details in **Fig S7**).

The strain on $WSe_2$ and $WS_2$ layers is estimated to be 1 ± 0.5 % and 2 ± 0.25 % (**Fig 2f, Fig S7**), respectively. These values are in agreement with strain parameters expected in transition metal dichalcogenides with nanobubbles[44]. The external tensile strain reduces the lattice parameter difference between $WS_2$ and $WSe_2$ and thereby increases the moiré periodicity. Additionally, the moiré unit cell undergoes an angular distortion (**Fig S8**) of 5-7% with different moiré diagonal lengths, suggesting the presence of heterogeneous shear and axial strains. Finally, we compared the effects of external strain on reconstruction with a similar analysis on the regions without external strain where the extent of reconstruction was 2.5 ± 0.4 (**Fig S20c**). The external strain reduces the reconstruction by 19%.

To observe the effects of structure on the electronic states, we perform cryogenic hyperspectral monochromated STEM-EELS, which measures the imaginary part of the dielectric constant, similar to optical reflectance. Spectra are acquired from similar regions as in **Fig 2** with a spectral resolution of ~10-15 meV and spatial resolution of approximately 1 Å (**Details in Fig. S14**).

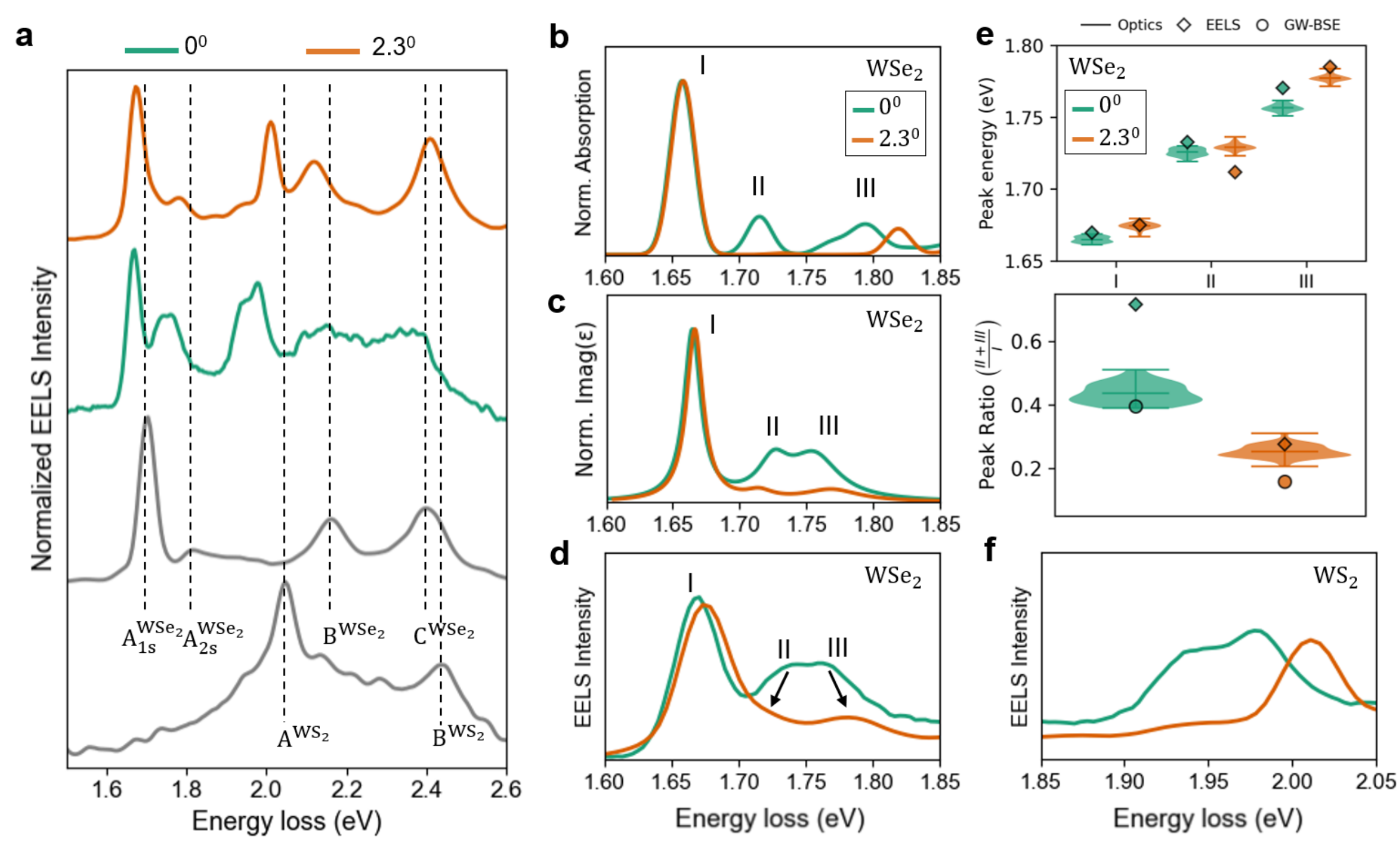

***Figure 3: Evidence of structural reconstruction on the excitons in $WSe_2/WS_2$.*** *(a) Comparison of monochromated EEL spectra from the reconstructed (green) and partially-reconstructed (orange) $WSe_2/WS_2$ heterostructures, monolayer* $WSe_2$ (gray), *and monolayer $WS_2$* (gray)*. The energy resolution is 11, 14, and 25 meV for reconstructed (green), partially-reconstructed (orange), and monolayer regions, respectively. (b) Optical absorbance spectrum calculated via GW-BSE, (c) Imaginary part of the dielectric function extracted from optical reflectance and (d) monochromated EEL spectra of $WSe_2$ excitons in the reconstructed (green) and partially-reconstructed (orange) heterostructure showing three exciton peaks labelled as peaks I-III. (e) Statistics of peak positions and peak ratio $\frac{II+III}{I}$ extracted from dielectric response fitting of large-area mapping of optical reflection contrast across the 0° and 2.3° heterostructures. The diamond and circle markers represent the corresponding deconvoluted peaks from EEL and GW-BSE spectra, respectively. The shaded bands represent deviations in the peak positions and the peak ratios (f) EEL spectra from the reconstructed (green), and partially-reconstructed heterostructures (orange) showing the $WS_2$ excitons.*

The representative monochromated EEL spectra from heterostructures and monolayer regions is shown in **Fig 3a**. The $WSe_2$ monolayer has four main exciton peak features at 1.7 eV ($A_{1s}$), 1.8 eV ($A_{2s}$), 2.2 eV (B), and 2.4 eV (C). $WS_2$ has two major features at 2.1 eV ($A_{1s}$) and 2.5 eV (B). The heterostructure formation leads to a red-shift of the $A_{1s}$ in comparison to the monolayers, corroborating with similar reports from optical reflectance[13,22,29,30]. The exact red shift depends on the structural reconstruction in the moiré pattern and external strain, which are explained in detail below.

The structural reconstruction gives rise to multiple $WSe_2$ exciton peaks (between 1.6-1.8 eV), labelled as peaks I-III. We limit our discussion to only the $WSe_2$ layer because the $WS_2$ layer has only one peak (between 1.9-2 eV) which makes it difficult to assess the impact of structural reconstruction based on just spectral features. The EEL measurements were further compared with ab initio GW plus Bethe-Salpeter equation (GW-BSE) calculations[13,22,30] and optical reflectance measurements (**Fig 3b-d, S12**, Details about methods in supplemental).

In all three methods, $A_{1s}$ is split into three peaks according to GW-BSE (**Fig 3b**), optical reflectance (**Fig 3c**) and EELS (**Fig 3d**) measurements. However, the change in the structural reconstruction affects the relative peak I, II, and III intensities in the reconstructed and partially-reconstructed heterostructures. We quantify this variation by calculating the peak ratio, defined as the intensity of peak II+III/ intensity of peak I. The peak ratio of the reconstructed heterostructure is higher than in the partially-reconstructed case, strongly indicating a direct connection with the change in the structural reconstruction in **Fig 2**.

It is to be noted that there are subtle differences (~0.02 eV) in the peak positions and peak (II+III)/I ratio (**Fig 3e**) between GW-BSE and optical reflectance, strongly corroborating the slight differences in the area ratio we observe in **Fig 2d** and overall sample variability (**Fig 3e**). Similar subtle differences (~ 5meV) **(Fig 3c, d)** are also observed between EELS and optical reflectance due to temperature differences between the two measurements. We note that even though EELS spectral features (**Fig 3d**) are slightly broadened compared to optical reflectance (**Fig 3c**) and GW-BSE, these are primarily due to the spectrometer limitations and not any sample heterogeneities or electron beam damage (**Fig S13**).

The external strain is expected to cause additional changes in the exciton peaks. Experimentally, the effect of external strain on the $WSe_2$ layer cannot be easily accessed without spatial mapping. However, since reconstructed $WS_2$ only has one peak, an additional red shift is strongly attributed to external strain.  We observe an additional large redshift of ~60 meV in the broad $WS_2$ $A_{1s}$ exciton peak in EELS of the reconstructed heterostructure (**Fig 3f**), attributed to approximately 2% extrinsic tensile strain acting on the $WS_2$ layers (**Fig 2f)** and corroborating previous studies on optical reflectance measurements on $WS_2$[45].

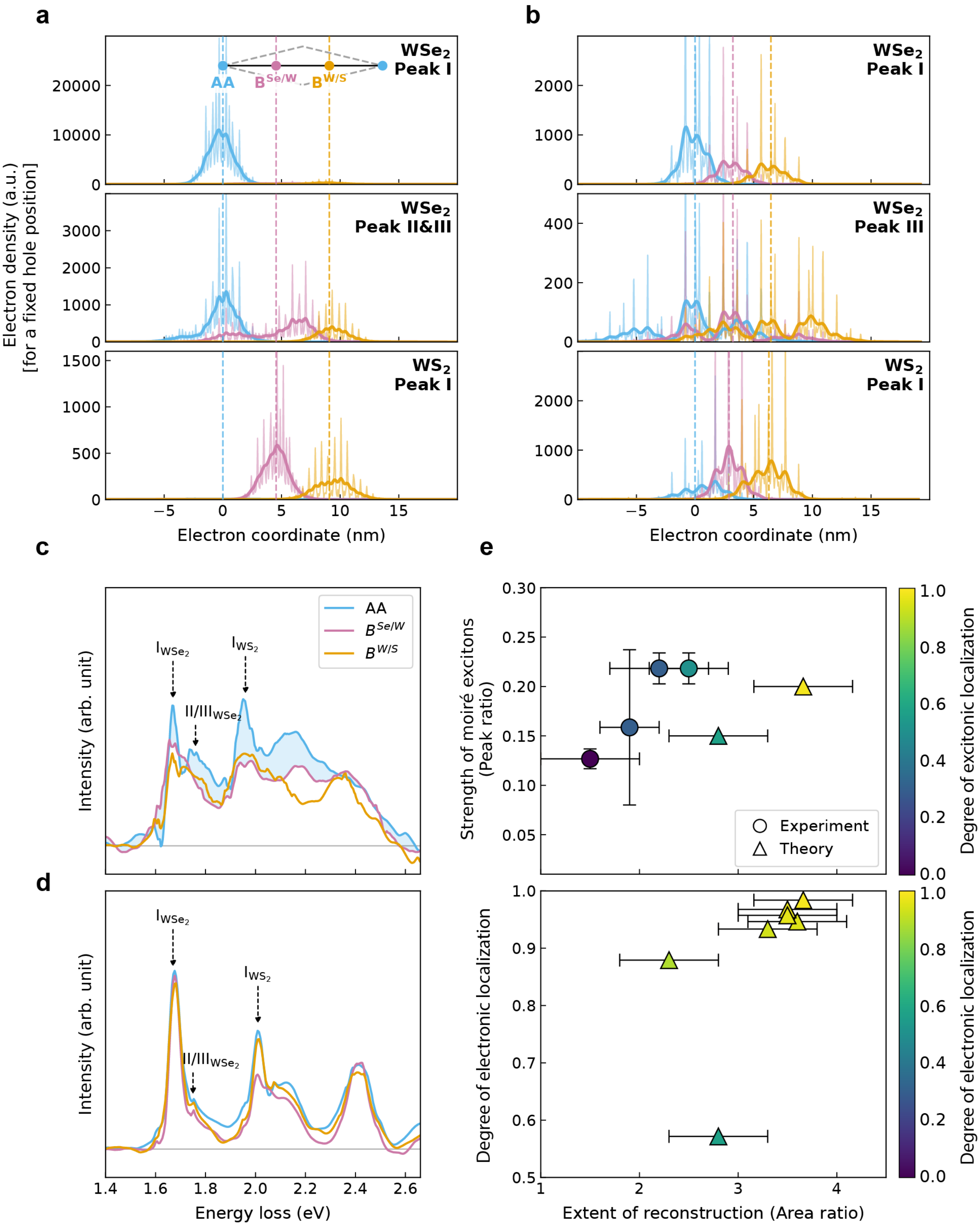

***Figure 4: Stacking site-specific exciton spectra in $WSe_2/WS_2$.*** *(a) Calculated photoexcited electron charge density of the exciton, plotted for fixed hole positions in the moiré superlattice. The hole is fixed at the high-symmetry AA, $B^{Se/W}$, and $B^{W/S}$ stackings, as shown in the inset. The exciton maps are plotted for the various peaks in the (a) reconstructed and (b) partially-reconstructed heterostructures. The solid lines are a guide to the eye, highlighting the electron*

*density spatial distribution. Site-specific averaged and background-subtracted STEM-EEL spectra from (c) reconstructed and (d) partially-reconstructed heterostructures. (e)* ***(Top)*** *Phase plot indicating the conditions for $WSe_2$ exciton localization at the AA sites. Solid circles show the four experimental data points, with the extent of reconstruction (area ratio) extracted from HAADF-STEM and the peak ratio $\frac{II+III}{I}$ extracted from optical reflectance. The color on the plot signifies the extent of exciton localization extracted from the relative exciton ratio ($\frac{I_{AA}-I_B}{I_{AA}}$) at the AA and Bernal stacking sites from STEM-EELS measurements (experiments) and the GW-BSE exciton calculations (theory).* ***(Bottom)*** *Phase plot showing the extent of electronic localization at the AA stacking as a function of the extent of structural reconstruction, extracted from the charge density of the reconstructed $WSe_2$ CBM.*

Besides the changes in spectral features, a varying extent of structural reconstruction will directly affect the extent of site-specific exciton localization. The theoretical exciton maps (**Fig 4 a,b**) were obtained from our GW-BSE calculations using the PUMP approach. The calculated excitonic states corresponding to the peaks in **Fig 3b** are given by $\psi^S(\mathbf{r_e}, \mathbf{r_h})$, where S is the excitonic state index, $\mathbf{r_e}$ and $\mathbf{r_h}$ are electron and hole coordinates, respectively. To map the excitons in the moiré superlattice, we fix the photoexcited hole position, $\mathbf{r_h}$, at a high-symmetry stacking site, and plot the electron density: $|\psi^S(\mathbf{r_e}, \mathbf{r_h})|^2$ (more details in **Fig S22-23**). Experimentally, we performed site-specific averaging of STEM-EELS datasets from about 10 moiré stacking sites, equating to about 2.5 moiré unit cells (**Fig S15-16, details about masking in Fig S21**). All the unit cells had similar elongated moiré unit cells as shown in **Fig 2**. The background-subtracted site-specific spectra are shown in **Fig 4c, d.** The details of the background subtraction are presented in **Fig S17-19**.

In the reconstructed lattice without any external strain, we computationally observe that all three peaks in $WSe_2$ (**Fig 3b**) have maximum amplitude at the AA sites (**Fig 4a**). Peak I, in particular, exhibits strong localization: the photoexcited electron density is significant only when the hole is fixed at the AA stacking site (**Fig S25**). This is in good agreement with previous calculations and experimental findings[13,22]. From a theoretical standpoint, excitons corresponding to peaks II and III also exhibit maximum electron density when the hole is fixed at the AA stacking site, while weak electron-density modulations appear when the hole is fixed at other stacking sites, corroborating the experimental findings of the maximum intensity contrast at the AA site. In contrast, the $WS_2$ exciton peaks are found to be localized at the Bernal stacking sites.

The presence of external strain changes the localization landscape. Experimentally, we observe that in the presence of 1% external tensile strain on $WSe_2$. Both peak I and II+III are localized at the AA sites (**Fig 4c**) which can be visualized by the blue shaded regions that represent the additional peak intensities at the AA sites, corroborating with theoretical expectations as well (**See Fig S24**). However, there are also significant intensities of peak I-III at the Bernal sites, which suggests that there is a lower extent of exciton localization. The intensity at the Bernal sites reduces slightly (by approximately 15%) in the absence of external strain (**Fig S20**). This discrepancy can be correlated to the difference in the $B^{Se/W}$ or $B^{W/S}$ / AA area ratio in **Fig 2**. In the theoretical case, this area is 2.8 ± 0.9 whereas in the experimental scenario, the observed values are approximately 2.2 and 2.5 in the strained and unstrained conditions, respectively. In the partially-reconstructed case, we do not observe any significant localization either in the GW-BSE calculations or experiments despite the presence of three exciton peaks I-III, which are considered hallmarks of exciton localization[12].

Based on these results, we have a generalized framework where the degree of exciton localization is defined. We use four metrics to create this framework. 1) Extent of reconstruction which signifies how much the AA and Bernal sites contract/expand relative to each other. We already captured this parameter using the HAADF-STEM images in **Fig 2d**. This is shown as the x-axis in **Fig 4e**. 2) Strength of moiré excitons which signifies how structural reconstructions lead to a change in electronic structure and influence optical selection rules, leading to multiple exciton peaks; I-III. Experimentally, this parameter is derived from optical reflectance measurements defined as the ratio of the intensity of peak II+III/ intensity of peak I (**See Fig 3e**) . For cases where there are strain heterogeneities, the same peak ratio can be used since optical reflectance provides an average measurement of strained and unstrained regions[46] 3) The degree of electronic localization obtained by comparing the localization around the AA and Bernal (W/Se) sites. We define the metric as $C = \frac{I_{AA} - I_{B^{W/Se}}}{I_{AA}}$ where C approaches 1.0 for strong AA localization and zero for delocalized states. The latter metric is a high-throughput method to understand localization of states via density functional theory band structures of multiple twist angles. Strong electron localization is a necessary condition for excitonic localization. We cannot determine this parameter experimentally. This is hence shown in the bottom plot of Fig **4e (Details in Supplemental Note 5.6 and Fig S26,27).** 4) Degree of exciton localization which is defined mathematically as $(\frac{I_{AA} - I_B}{I_{AA}}) \times 100$. This is experimentally extracted from spatially resolved STEM-EELS measurements and computationally extracted from GW-BSE exciton calculations. It is shown by the color bar in the phase plot of extent of reconstruction vs strength of moiré excitons (**Fig 4e**). These parameters together form the phase plot where the degree of exciton/electronic localization is defined by the color bar. Importantly, these metrics retain their validity under external strain, as strain-induced lattice modifications directly alter the relative ratios of $B^{Se/W}$or $B^{W/S}$ / AA stacking domains within the moiré unit cell.

On the other hand, the $WS_2$ excitons under ideal conditions are weakly localized at the Bernal stacking sites. However, in the presence of external strain, we observe a strong localization of $WS_2$ excitons at the AA sites. There is also an appearance of a distinct peak at 2.3 eV at Bernal stacking sites. The theoretical interpretation of this peak is computationally demanding and warrants dedicated future investigation beyond the scope of the present work. In the partially-reconstructed lattice, the $WS_2$ exciton appears weakly localized at AA and $B^{W/S}$ sites.

## Conclusions and Outlook

Taken together, our results establish a mechanistic framework for understanding how structural reconstruction, external strain, and excitonic spectral signatures jointly determine whether excitons are localized, delocalized, or displaced from their theoretically expected stacking sites. Critically, we demonstrate that the appearance of multiple excitonic resonances in optical spectra is not a sufficient indicator of real-space exciton localization. This necessitates revision of how moiré exciton experiments are interpreted.

We identify several quantitative metrics that reliably govern exciton confinement, 1) the extent of reconstruction, 2) peak ratio, 3) the degree of electronic localization and 4) the degree of exciton localization. These metrics may be extended beyond the bilayer $WSe_2$/ $WS_2$ systems to other TMD bilayers, trilayers, and magnetic semiconductor moiré systems. The nuances of the area ratio parameter and peak ratio defined will vary on a case-by-case basis. Overall, the phase plot provides a robust framework for distinguishing true moiré exciton localization from spectrally similar but spatially delocalized states, even in partially-reconstructed and lattices with heterogeneous strain.

This work also has a few limitations. First, the moiré heterostructure shown in **Fig 2** shows rotational symmetry breaking due to anisotropic and asymmetric strain on each layer. Computationally, it is difficult to conclude what impact it has on excitons. We do observe some unidentified peaks at higher energies in EELS, but the corresponding structural resolution is limited which can be improved with longer stability holders and better sample designs for sustaining high electron dose. Second, the reflectance measurements are performed at 4K whereas STEM-EELS is performed at 100 K. The temperature can itself cause some delocalization effects. The STEM-EELS experiments using the liquid He holders[47] can provide an accurate answers to this question.

Establishing reliable optical signatures of microscopic structure and exciton localization is crucial for advancing the broader field of moiré materials. Raman and photoluminescence spectroscopy, for example, provide practical fingerprints of layer thickness, strain, and defect density in two-dimensional materials. Similarly, optical exciton spectra could offer a scalable means of identifying localization regimes that otherwise require demanding nanoscale measurements. Our results identify an important limitation of such optical assignments. They establish structural reconstruction and the associated strain landscape, rather than simply the number of optical resonances, as the key factors governing localization. The resulting phase plot provides mechanistic guidance for identifying and engineering localized excitons in moiré heterostructures. This is directly relevant to emerging strain-engineering and straintronics approaches[47,48]. The present work also extends the analysis to excitons in the $WS_2$ layer. Our measurements and calculations show that their real-space localization can nevertheless evolve substantially with reconstruction and strain. This finding demonstrates that neither the appearance of multiple peaks nor the persistence of a single peak provides a complete measure of exciton localization.

Overall, the ability to drive exciton localization and site hopping through local structural modulation opens pathways toward reconfigurable excitonic lattices and strain-addressable quantum optoelectronic devices. More broadly, the correlative structural–spectral methodology demonstrated here offers a general framework for probing exciton behavior in realistic quantum materials, where lattice relaxation and strain [1,50–54] are unavoidable and fundamentally shape emergent electronic and excitonic properties.[1,50–54]

## Acknowledgements:

Support for sample fabrication, second-harmonic generation spectroscopy, and cryogenic optical microspectroscopy was provided by the U.S. Department of Energy, Office of Science, Basic Energy Sciences in Quantum Information Science under award number DE-SC0022289. Electron microscopy research conducted as part of a user project at the Center for Nanophase Materials Sciences (CNMS), which is a US Department of Energy, Office of Science User Facility at Oak Ridge National Laboratory. Work at the Molecular Foundry was performed under user projects and supported by the Office of Science, Office of Basic Energy Sciences, of the U.S. Department of Energy under Contract No. DE-AC02-05CH11231. The authors also acknowledge the use of facilities within the John M. Cowley Center for High Resolution Electron Microscopy at Arizona State University. Sandhya Susarla would like to acknowledge ASU start-up support and Army Research Office Early Career Program under Award Number W911NF-26-1-A047. The TEM efforts at ASU were partially supported by NSF CAREER program. Sriram Sankar would like to acknowledge the support of the Fulton Fellowship, Army Research Office and NSF CAREER program. The theoretical calculations were funded by the Center for Dynamics and Control of Materials (CDCM), which is supported by the National Science Foundation under NSF Award Number DMR-2308817. Rwik Dutta acknowledges support from the Texas Quantum Institute graduate fellowship. We acknowledge the Texas Advanced Computing Center (TACC) at The University of Texas at Austin for providing computational resources to carry out the calculations. This work used resources from Stampede3 through allocation PHY250206 from the Advanced Cyberinfrastructure Coordination Ecosystem: Services and Support (ACCESS) program, supported by NSF grants 2138259, 2138286, 2138307, 2137603, and 2138296. Seth Tongay acknowledges support from NSF CBET 2330110 and Lawrence Semiconductors. K.W. and T.T. acknowledge support from the JSPS KAKENHI (Grant Numbers 21H05233 and 23H02052) , the CREST (JPMJCR24A5), JST and World Premier International Research Center Initiative (WPI), MEXT, Japan.

**Author contributions:**
*Project conception*: AR, SS *Project supervision*: AR, SS, MHN; *Sample fabrication*: MD, DB; *ADF-STEM-EELS experiments*: JH, SS; *Optical spectroscopy*: MD; *Experimental data analysis*: SS, MD (Equal contribution); *DFT* and *GW-BSE calculations*: GE, RD, JRC, MHN; *Construction of atomic structures for STEM simulations*: GE, SS; *Participation in Scientific Discussion*: P.E., P.J.H, S.T, J.R.C; *hBN Crystals*: T.T, K.W *Writing*: MD, SS, GE, SS with inputs from all the authors. *Editing:* AR, MHN, MD, SS, SS with inputs from all authors.

**Supplemental Information:** The supplemental information can be requested from corresponding authors upon reasonable request.